\documentclass[12pt]{article}
\begin{document}
\begin{center}
{\large Effects of the Inverse Square Potential on Nucleon-Nucleon Elastic Scattering in the Bethe-Salpeter Equation}
\end{center}
\begin{center}
{Susumu Kinpara}
\end{center}
\begin{center}
{\it National Institute of Radiological Sciences, Chiba 263-8555, Japan}
\end{center}
\begin{abstract}
Bethe-Salpeter equation is applied to nucleon-nucleon elastic scattering at the intermediate energy.
The differential cross section and the polarization are calculated in terms of the phase shift analysis method
using the two-body potential derived from the Bethe-Salpeter equation.
The lowest-order Born approximation for the K-matrix is corrected by including the inverse square part
of the potential.
\end{abstract}
\hspace*{4.mm}
One of the main purpose for study of nuclei is to understand the nuclear force between nucleons
and the effects on the nuclear many-body system.
The meson exchange picture and the two-body interaction play a decisive role to describe the nuclear structure.
To give an example the higher-order diagrams beyond the Hartree or the Hartree-Fock approximation elucidate why
the lowest-order treatments of the proper self-energy reproduce observables of the nuclear system$\cite{Kinpara}$.
And the nuclear effective interactions as illustrated by the pairing interaction in the nonrelativistic formulation are largely connected with the two-body correlations in nuclei.
\\\hspace*{4.mm}
Strictly the meson exchange potential used to calculate founds on the screened coulomb type, thus,
the relation with two-body interactions in free space is still an open question.
In order to treat the relativistic system composed of two nucleons the Bethe-Salpeter (BS) equation 
is indispensable to the underlying framework of the theory
and the validity on the field theoretical points of view has been reported in the literature for a long period.
The geometry of the space-time is four dimensional and the interaction contains relativistic effects
such as the retardation effect, which is often neglected to make the calculation tractable 
in a lot of models of the potential.
\\\hspace*{4.mm}
The integral equation on the two-particle Green function is used to derive the BS equation.
By equating the pole terms on the both sides 
the differential equation is derived for the bound state composed of two nucleons, that is, deuteron.
The singular potential is necessary to evaluate the quantity
and the BS equation is expected to provide analyses at the short-range or the high momentum transfer region.
Recently it has been shown that the binding energy and the electric quadrupole moment of deuteron 
are obtained by solving the ladder approximated BS equation with the asymptotic approximation$\cite{Kinpara2}$.
It is the main subject of the present study to investigate scattering states between two nucleons, for which
the boundary condition exists in the asymptotic region and determines the observables.
\\\hspace*{4.mm}
The BS amplitude is expanded by a set of the Gamma matrices $\Gamma_i$ and which results in
the simultaneous equations on the expansion coefficients.
They act as the wave functions $\Psi_i$ ($i$=$S$,$V$,$PS$,$AV$,$T$), in which 
each subscript denotes the scalar, vector, pseudoscalar, axial-vector and anti-symmetric tensor sectors respectively.
By the Fourier transform they are converted from the four dimensional momentum space to the configuration space ones
so as to proceed calculations for the present study.
\\\hspace*{4.mm}
When terms of the total energy are taken into account exactly the equation of the vector sector $\Psi_V$ 
is independent of the other components provided that the auxiliary condition $p_0\Psi_0(p)=0$
is imposed on the zeroth vector component $\Psi_0(p)$, where $p_0$ is the zeroth component 
of the relative four-momentum in the center of mass system.
Calculating the binding energy of deuteron the vector component is removed 
because the zeroth component is assigned the spin $S$=0.
The three polar components of the anti-symmetric tensor are substituted to construct the $S$=1 bound state of deuteron.
\\\hspace*{4.mm}
Doing the Gamma matrix expansion and applying the auxiliary condition to the origin of the relative time ($t$=0) in the configuration space the equation results in the form equivalent to the Schr$\rm{\ddot o}$dinger eigenvalue equation as
\begin{equation}
[-\delta_{ij}\frac{1}{2 \mu}\mbox{\boldmath $\nabla$}^2+V_{ji}(\mbox{\boldmath $x$})]\Psi_j(\mbox{\boldmath $x$})
=\frac{k^2}{2 \mu} \Psi_i(\mbox{\boldmath $x$}),
\end{equation}
\begin{equation}
k = \sqrt{\mu E},
\end{equation} 
where $\mu$ and $\it E$ denote the reduced mass of two nucleons 
and the laboratory energy of the incident nucleon respectively.
The dummy index is applicable to $j$.
The subscripts $\it i,j$=$V$, $AV$ or $PS$ on $S$=0 and $\it i,j$=$\it x,y,z$ the cartesian coordinate 
of the polar-tensor on $S$=1.
When $S$=0 the potential is assumed to be $V_{ji}(\mbox{\boldmath $x$})=\delta_{ji}V_{i}(\mbox{\boldmath $x$})$.
It has been reported on the study of the electric quadrupole moment of deuteron 
that the mixing term in the polar-tensor equation nearly cancelled out
the joint effects of the axial-vector and the pseudoscalar components, thus, it is neglected in the 
following calculation for $S$=1 part as well as that of the pseudoscalar equation applied for $S$=0
in the numerical calculations.
\\\hspace*{4.mm}
It is remarkable that Eq. (1) is derived without the nonrelativistic approximation 
as the expansion by $v/c$, then, $k^2$ may take any positive value of the incident energy for scattering states.
We make a suggestion that the pseudoscalar component $\Psi_{PS}(\mbox{\boldmath $x$})$ 
is used for the singlet ($S$=0) state to construct 
the proton-neutron ($p$-$n$) elastic scattering along with a set of $\Psi_T^{0i}(\mbox{\boldmath $x$})$ ($i=x,y,z$) 
for the triplet ($S$=1) state which has also been applied to describe the bound state quantities.
Since the mixing term in the pseudoscalar equation is dropped out these two spin states $S$=0 and $S$=1 do not couple 
mutually similar to the two-body interaction potential $V_i(\mbox{\boldmath $x$})$.
\\\hspace*{4.mm}
When one investigates the scattering problem the partial wave analysis method is generally used for proton-proton ($p$-$p$)
case and formulated by the M-matrix with the isospin $T$=$T_3$=1$\cite{Stapp}$.
This formulation is extended to $p$-$n$ elastic scattering, in which both of the isosinglet ($T$=0) $M^0$ and
the isotriplet ($T$=1) $M^1$ matrices are prepared to obtain the resulting observables.
There is reason why the zeroth vector component is not adapted for $S$=0, that is, the spin function $\Gamma_V$ 
is symmetric ${}^t\Gamma_V=\Gamma_V$, then, the parity $(-1)^L=+(-)1$ are assigned $T=0(1)$ states accordingly.
On the other hand, the spin $S$=1 is carried by $\Gamma_T$ of the tensor components
which is symmetric ${}^t\Gamma_T=\Gamma_T$ and the assignment of the parity is equal to that in the usual formulation.
\\\hspace*{4.mm}
For the mixing between $S$=0 and $S$=1 is not assumed as stated above, the partial waves are decomposed into 
three parts, ($S$,$L$,$J$)=(0,$L$,$L$), (1,$L$,$L$) and (1,$J$$\mp$1,$J$).
The last part allows mixing between two states $L$=$J$$\mp$1 and which is represented by 
three real parameters $\delta_{J\mp1 J}$ and $\epsilon_J$.
The nuclear bar phase shift method is used to express them$\cite{Stapp}$.
The mixing parameters $\epsilon_J$ play a decisive role in giving suitable magnitudes 
on the M-matrix elements $M_{10}$ and $M_{01}$ by which the value of the polarization may be influenced much.
\\\hspace*{4.mm}
In order to determine the phase shift parameters for $p$-$n$ elastic scattering the formulas
\begin{equation}
\rm{tan}\;\delta_{\it L} 
= -2{\it k \mu} \int_0^\infty {\it r}^2 d{\it r}\; {\it j}_{\it L}({\it k}{\it r}) {\it V_i}({\it r}) 
\psi_{\it L}({\it k}{\it r}),
\end{equation}
are applied with use of the potential derived from the BS equation in $S$=0 case.
Here, $j_L(kr)$ and $\psi_L(kr)$ are the spherical Bessel function of the order $L$ 
and the exact solution under the potential in Eq. (1) respectively.
The $S$=1 case is analogous to Eq. (3).
It corresponds to the Born approximation in the K-matrix formulation
when $j_L(kr)$ replaces $\psi_L(kr)$ by the lowest-order approximation on the strength of the potential $V_i(\it r)$.
The K-matrix is 2$\times$2 real symmetric matrix and related to the phase shift parameters 
of the S-matrix represented by the equation $S=(1+iK)^{-1}(1-iK)$ in the algebraic notation.
\\\hspace*{4.mm}
In the present meson exchange model the isoscalar scalar meson $\sigma$, the isoscalar vector meson $\omega$,
the isovector pseudoscalar meson $\pi$ and the isovector vector meson $\rho$ are taken into account. 
Values of the coupling constants and masses are determined appropriately referring to calculations for 
the relativistic nuclear many-body system and the structure of deuteron$\cite{Kinpara2}$.
The electromagnetic interaction is turned off in the present study which has an influence on
the results particularly at the forward direction of the center of mass system ($\theta_c\leq{\rm 10^\circ}$) 
in $p$-$p$ system.
\\\hspace*{4.mm}
Concerning the interaction of pion the pseudovector coupling is employed and the propagator is multiplied by
the four dimensional cut-off function $\Lambda^{\rm 2}/(\Lambda^{\rm 2}-p^{\rm 2})$ in the momentum space.
By means of the cut-off procedure the inverse fourth power potential shape ($\sim r^{\rm -4}$) at $r \rightarrow 0$ is modified to the inverse square potential one ($\sim r^{\rm -2}$) and which
makes calculations for the phase shifts (Eq. (3)) feasible.
Another important feature of the pseudovector coupling of the pion interaction 
is the property of the angular momentum changing ($L\rightarrow L \pm 2$) in the two-body potential 
which effectively acts like the tensor force giving a suitable size of the mixing parameter 
unlike the case of the pseudoscalar coupling interaction.
\\\hspace*{4.mm}
While the Born approximation works well for the Coulomb potential, the nuclear potential is much stronger
at the short-range region and for the exact wave function $\psi_L(kr)$ we need to take into account
the inverse square potential originating in the Feynman propagator therein.
Instead of the exact solution by numerical procedure only the leading-order inverse square part of the potential
is left and the approximate solution $\psi_L(kr,g)$ is substituted for $\psi_L(kr)$ in the present study.
For scattering states Eq. (1) under the inverse square potential 
is solved analytically and it is given as
\begin{equation}
\it{\psi_L(kr,g)}=\rm{sec}(\frac{\it{L}+\frac{\rm 1}{\rm 2}-\nu}{\rm 2}\pi)
\sqrt{\frac{\pi}{\rm 2 \it{kr}}}\it{J_\nu}(kr).
\end{equation}
The order of the Bessel function $\nu \equiv \sqrt{( L+ \frac{1}{2} )^2 - g}$ is dependent on the strength $g$ of 
the inverse square potential part ($V$($r$) $\sim$ -$g\;M^{\rm -1}\;r^{\rm -2}$), where $M$ is the nucleon mass. 
When we solve the equation the Neumann function part is dropped by assuming the phase shift
$\delta_L=(L+1/2-\nu)\pi/2$ on the inverse square potential.
Thus, the elements of the K-matrix is improved by the approximate function $\psi_L(kr,g)$ in this manner.
It is verified that the spin $S$=0 part is not influenced much by the procedure  
so the correction is not done for the Born term.
Concerning the spin $S$=1 part we correct only for ($L$,$J$)=(0,1) state as to the $p$-$n$ scattering case 
to enhance the phase shift parameters of the $J$=1 part in the K-matrix suitably.
\\\hspace*{4.mm}
Instead of the direct expansion 
of the K-matrix Eq. (3) is used for the (0,1) element by multiplying the factor $F(\mit\Lambda)$ to the Born term as
\begin{equation}
<L^\prime \mid V \mid \nu>\;=\;<L^\prime \mid V \mid L> F(\mit\Lambda),
\end{equation}
\begin{equation}
F(\mit\Lambda) \equiv \; <L^\prime \mid V(\mit\Lambda) \mid \nu>/<L^\prime \mid V(\mit\Lambda) \mid L>.
\end{equation}
Here, the bracket notation is used to clarify some points on the cut-off $\mit\Lambda$ 
for the propagator of pion explained previously.
The argument $\mit\Lambda$ denotes that the cut-off $\mit\Lambda$ in the potential is allowed to vary as a 
parameter from $\mit\Lambda_0$ to the infinity $\infty$.
On the other hand, the fixed cut-off $\mit\Lambda_0$ is in the potential $V$ and $\nu$ to make the integral of the Born term be convergent.
The value of $\mit\Lambda_0$$\;\sim\;$500\,-\,600$\;$MeV is tentatively used by
taking account of the calculation for the binding energy of deuteron$\cite{Kinpara2}$.
As a result of the numerical calculation it is shown that $F(\mit\Lambda)$ remaines finite as $\mit\Lambda$$\,\rightarrow\,$$\infty$
and the results with $F(\infty)$ give remarkable effects on the scattering phenomena compared with $F(\mit\Lambda_0)$.
So, $F(\infty)$ is chosen to calculate the higher-order corrections in the present study.
\\\hspace*{4.mm}
The differential cross section for $p$-$n$ elastic scattering is given as
\begin{equation}
\frac{d \sigma_{p-n}}{d \Omega}=\frac{1}{8}\sum_{T=0,1}{\rm Tr}\,M^TM^T{}^\dagger,
\end{equation}
where $M^0$ and $M^1$ denote the M-matrices of the isosinglet and isotriplet respectively.
In Fig. 1 the angular dependence of the differential cross section at the laboratory energy $E\,$=$\,$310$\,$MeV 
is shown in two cases (V)the lowest-order approximation in the K-matrix theory
and (K)inclusion of effects of the higher-order terms by the procedure stated above.
By including the corrections the curve becomes deeper at the center of mass scattering angle $\theta_c=90^\circ$ 
and shows the desirable V-shape.
Thus, the short-range character of the nuclear force is responsible for the angular dependence largely
and in fact contributes to determine the incident energy dependence of the $p$-$n$ scattering favorably.
\\\hspace*{4.mm}
The polarization is another means of investigating properties of the nuclear force 
and the accurate phase shift parameters with the differential cross section equally.
The results of the calculation for polarization are shown in Fig. 2 
where the signs (V) and (K) denote same as those in Fig. 1.
Since the present formulation assumes the identical two fermions in the isospin space
the curve is symmetric at $\theta_c=90^\circ$ different from the asymmetry seen in the experiments.
The result of the lowest-order calculation (V) is not satisfying for lack of the sufficient tensor force.
Then, by including the higher-order effects the curve turns the sign successfully.
The higher-order calculation results in an underestimate particularly at the forward direction ($\theta_c < 90^\circ$)
compared with the experiment in which the maximum value reaches 0.4 at around $\theta_c = 30^\circ$$\cite{Cheng}$.
The discrepancy may be attributed to the approximate form of $\psi_L(kr,g)$ adopted 
for calculating the K-matrix, which would be improved by performing calculations next to leading order perturbatively
or correcting the other ($L$,$J$) states in addition to (0,1) state.
\\\hspace*{4.mm}
The BS formalism is also applicable to $p$-$p$ elastic scattering 
by adding the inhomogeneous term to the BS equation$\cite{Salpeter}$.
While the role of the various components in the BS amplitude is not necessarily confirmed, 
the pseudoscalar component is assigned the $S$=0 scattering state as well as the $p$-$n$ case.
The S-wave ($L$=0) is essential for $p$-$p$ elastic scattering to interpret the observed straight line shape 
at $\theta_c > 10^\circ$ of the differential cross section.
As to the $S$=1 state a set of the polar-tensor components is appropriate following the $p$-$n$ case.
\\\hspace*{4.mm}
The differential cross section for $p$-$p$ elastic scattering is given 
by changing the sum over $T$ in Eq. (7) as $\sum_{T=0,1} \rightarrow 2 \sum_{T=1}$.
It is verified that the numerical calculation results in an overestimate of the experimental one 
when parameters of the coupling constants, the meson masses and the cut-off parameter of pion in the present model 
are same as those of the $p$-$n$ elastic scattering.
In order to make up for the ladder approximation
the pion-nucleon pseudovector coupling constant $f$ is reduced to $f\sim\,$0.05
from the standard value $f\approx\,$1$\cite{Lyder}$ at $E\,$=$\,$310$\,$MeV incident energy.
The suppression of the neutral $\pi^0$ meson exchange force indicates 
that the lowest-order ladder diagram in the irreducible kernel is not sufficient and prompts us to correct
by the higher-order multi pion processes largely.
\\\hspace*{4.mm}
The triplet $P$ states ($L$,$J$)=(1,0), (1,1) and (1,2) are fundamental to construct $p$-$p$ elastic scattering
and among them only (1,2) state is related to the mixing parameter $\epsilon_2$.
Then, the (1,2) element of the 2$\times$2 K-matrix is calculated by using $\psi_{\rm 1}(kr,g^\prime)$ 
instead of $j_{\rm 1}(kr)$, in which $g^\prime$ denotes the strength of the inverse square part of the potential.
One complicated issue in the present formulation is that the pion exchange force is attractive 
when $T$=1 in the $S$=1 state, therefore, the $g^\prime$ becomes too strong to leave the $\nu^\prime$ real even in $L$=1.
The imaginary part invalidates the phase shift method to describe the $p$-$p$ elastic scattering consequently.
Dealing with the situation we make a decision to move the parameter $g^\prime$ at $-\infty < g^\prime \leq \frac{1}{4}$, 
by which the results are made feasible to compare with the experiments 
maintaining the framework of the phase shift method.
\\\hspace*{4.mm}
It has been found that the required strong repulsive force is in $S$=$T$=0 of the $p$-$n$ system. 
The inverse square potential in the pseudoscalar equation is mainly given by the isoscalar $\sigma$ and $\omega$ mesons as
\begin{equation}
g^\prime=\frac{g_\sigma^2-4 g_\omega^2}{(2 \pi)^2}+\frac{3f^2\Lambda^2}{(2 \pi)^2m_\pi^2}.
\end{equation}
Here, $g_\sigma$ and $g_\omega$ denote the coupling constants accordingly.
Meanwhile, the leading order of the pion exchange force with the cut-off function is attractive 
and less important.
Ultimately the expected strong repulsive force arises from the $\omega$ meson exchange interaction.
\\\hspace*{4.mm}
The replacement for $g$ is interpreted as manifestation of $\pi^{+}$-$p$-$n$ three-body system 
in the intermediate state.  
Since the positive pion ($\pi^{+}$) carries the isospin $T$=1 
the intermediate $p$-$n$ state is recognized as $T$=0 provided that the transition from the initial $p$-$p$ state is done by a charge-independent interaction conserving the isospin.
By the anti-symmetric property of the pseudoscalar sector (${}^t\Gamma_{PS}=-\Gamma_{PS}$) giving the spin function on $S$=0 the orbital angular momentum of the $p$-$n$ system comes to $L$=1 and followed by $P$-wave ($L$=1) state of $\pi^{+}$ 
to construct the odd parity $J$=2 state as a whole.
\\\hspace*{4.mm}
The differential cross section for $p$-$p$ elastic scattering is shown in Fig. 3.
While the calculation (V) is done by the lowest-order approximation
the pion-nucleon pseudovector coupling constant $f$ is changed as the above mentioned way
to improve on the ladder approximated irreducible kernel effectively.
As seen in the calculation (K) the weakened $\pi^{0}$ exchange force 
is compensated by the higher-order correction which improves the shape of the curve well 
in comparison with the calculation (V) keeping the strong angular dependence under the Born approximation. 
The procedure is effective against the underestimate of the polarization that is larger
than the present calculation seen in Fig. 4[4,7].
Particularly, enhancement of the mixing parameter $\epsilon_{\rm 2}$ by the exact treatment of the K-matrix 
is probably significant to reproduce the polarization like the $p$-$n$ elastic scattering case.
\\\hspace*{4.mm}
In the $p$-$p$ system it has been found that the coupling constant of pion decreases as the incident energy of proton 
increases in order to reproduce the differential cross section of the elastic scattering.  
The energy dependence of the $p$-$p$ system is more obvious than the $p$-$n$ system in which there is little need to adjust
the parameter between the 10$\,$MeV and 200$\,$MeV.
The BS formalism is suitable to comprehend two-body nucleon-nucleon system for once the interacting lagrangian is set up
the higher-order calculations are carried out systematically about each component of the BS amplitude 
by various methods in the quantum mechanics.
Determining which one is appropriate to describe each channel of the scattering phenomena 
experimental facts on the spin observables are useful.
\\\hspace*{4.mm}

\newpage
${\bf Figure\;Captions}$\\\\
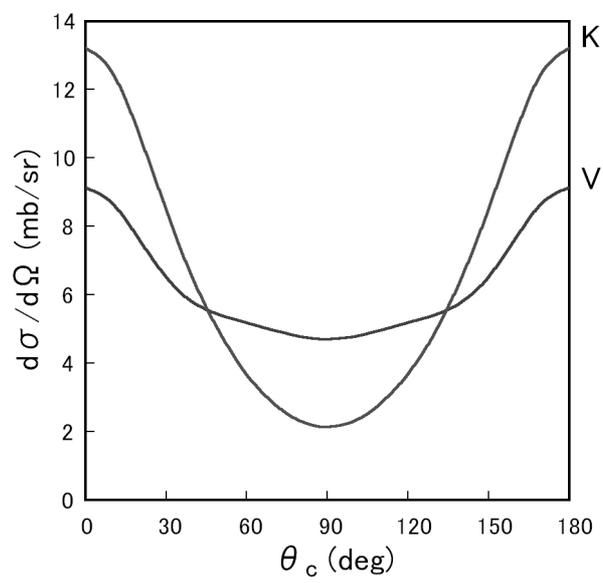
Figure 1: Proton-neutron differential cross section as a function of the center of mass scattering angle
at the laboratory energy of 310$\,$MeV.\\
V:The result of the calculation by the Born approximation.\\
K:Including the higher-order corrections.\\\\
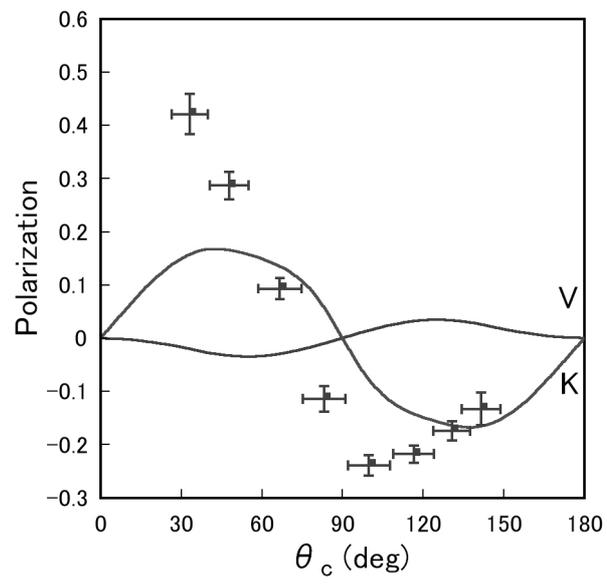
Figure 2: Proton-neutron polarization as a function of the center of mass scattering angle
at the laboratory energy of 310$\,$MeV.
The signs V and K denote same as those in Fig. 1.
The experimental data is from ref. $\cite{Cheng}$ for comparison.\\\\
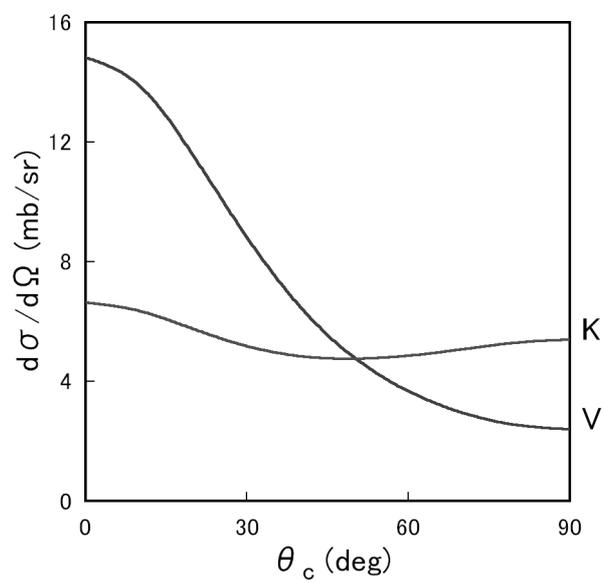
Figure 3: Proton-proton differential cross section as a function of the center of mass scattering angle
at the laboratory energy of 310$\,$MeV.\\
V:The result of the calculation by the Born approximation.\\
K:Including the higher-order corrections.\\\\
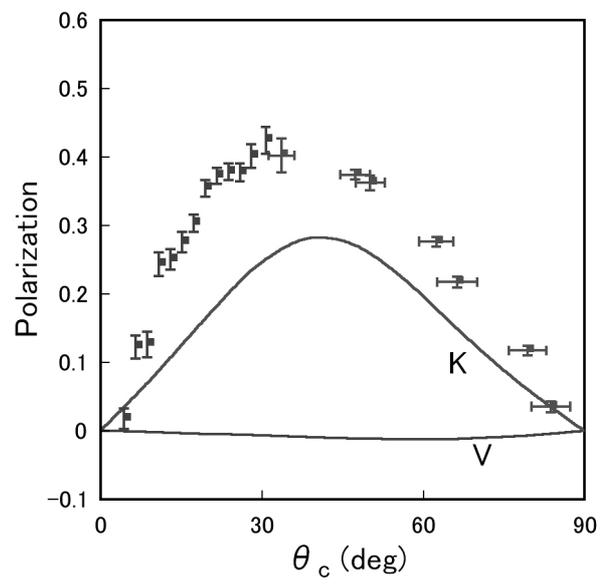
Figure 4: Proton-proton polarization as a function of the center of mass scattering angle
at the laboratory energy of 310$\,$MeV.
The signs V and K denote same as those in Fig. 3.
The experimental data are from ref. $\cite{Cheng}$ and from ref. $\cite{Besset}$ at 312$\,$MeV for comparison.
\end{document}